\documentclass[aps, showpacs,preprintnumbers,amsmath,amssymb,
               nofootinbib,twocolumn,superscriptaddress]{revtex4} 
\usepackage{graphicx}
\usepackage{verbatim}
\usepackage{slashed}
\usepackage{amsthm}
\usepackage{amsfonts}
\usepackage[dvips]{color}

\def\vev#1{\ensuremath{\langle #1\rangle}}

\def\beqn#1{\begin{equation}\label{#1}}
\def\eeqn{\end{equation}}

\def\beqna{\begin{eqnarray}}
\def\eeqna{\end{eqnarray}}

\def\beq{\begin{eqnarray}}
\def\eeq{\end{eqnarray}}
\def\bea{\begin{eqnarray}}
\def\eea{\end{eqnarray}}
\newcommand{\dm}{\text{\tiny DM}}

\def\ththMat#1#2#3#4#5#6#7#8#9{\ensuremath{\begin{pmatrix}#1&#2&#3\\
                                                          #4&#5&#6\\
                                                          #7&#8&#9\end{pmatrix}}}

\def\thoneMat#1#2#3{\ensuremath{\begin{pmatrix}#1\\
                                              #2\\
                                              #3\end{pmatrix}}}                                              
                                              
\def\order#1{\ensuremath{\mathcal{O}\left(#1\right)}}

\def\drate#1{\ensuremath{\Gamma(#1)}}

\begin{document}

\title{Inverse seesaw and dark matter in models with exotic lepton triplets}

\author{Sandy~S.~C.~Law}
\affiliation{Department of Physics, National Cheng-Kung University, Tainan 701, Taiwan}
\email{slaw@mail.ncku.edu.tw}

\author{Kristian~L.~McDonald}
\affiliation{Max-Planck-Institut f\"ur Kernphysik, Saupfercheckweg 1, 69117 Heidelberg, Germany}
\email{kristian@mpi-hd.mpg.de}

\date{11 April 2012}

\begin{abstract}
We show that models with exotic leptons transforming as $E\sim (1,3,-1)$ under the standard model gauge symmetry are well suited for generating neutrino mass via a radiative inverse seesaw. This approach realizes natural neutrino masses and allows multiple new states to appear at the TeV scale. The exotic leptons are therefore good candidates for new physics that can be probed at the LHC. Furthermore, remnant low-energy symmetries ensure a stable dark matter candidate, providing a link between dark matter and the origins of neutrino mass.
\end{abstract}

\maketitle


\section{Introduction}

The seesaw mechanism \cite{type1_seesaw} provides a simple means for generating naturally light standard model (SM) neutrinos. This simplicity, however, comes at a cost. If the Dirac Yukawa couplings in the neutrino sector are large [e.g.~\order{1}], as argued on the basis of naturalness, the gauge-singlet neutrinos must be very heavy, and there is essentially  no hope of experimentally producing them. Thus, if the seesaw is realized naturally in nature, the origin of neutrino mass will remain a mystery for the foreseeable future.  

Beyond the seesaw mechanism, however, there exist alternative approaches capable of explaining neutrino mass (see e.g.~\cite{Langacker:2011bi}). For example, light neutrinos can be realized naturally without invoking new high-energy physics if the right-handed neutrinos are low-energy composites of a confining hidden sector~\cite{ArkaniHamed:1998pf}.\footnote{Equivalently, via AdS/CFT, if they propagate in a hidden warped space~\cite{McDonald:2010jm}.}
 In principle, such models are easier to experimentally verify than a high-energy seesaw. In practice, models with light new physics can be similarly difficult to probe, due to very weak couplings. Indirect signals like flavor changing decays~\cite{Duerr:2011ks}, CMB signals~\cite{Okui:2004xn}, or galactic X-ray signals~\cite{Hundi:2011et}, often provide the best way to probe such scenarios. This is more promising than the somewhat bleak outlook afforded by the canonical seesaw, though it would be preferable if the new physics associated with neutrino mass was amenable to direct experimental production.

The likelihood of discovering new-physics related to neutrino mass increases if it includes new TeV-scale particles that are charged under the SM. Radiative models of neutrino mass provide a good example; here the SM couples to new fields that can potentially be probed at the LHC, while neutrino mass is generated as a small loop-effect~\cite{babu_zee}. These models are very promising from an experimental point of view, as the connection between small masses and new TeV-scale physics offers perhaps the best hope of discovery (see e.g.~\cite{Chen:2011de}).

With this in mind, it is worthwhile considering exotic states with \order{\text{TeV}} masses that have gauge invariant, renormalizable couplings to SM fermions. As one may expect, not many such candidates exist~\cite{DelNobile:2009st,Chua:2010me}, and most of those 
suitable for generating neutrino mass have been studied extensively. An interesting exception are the exotic leptons  $\smash{E_R=(E^0_R,\, E^-_R,\, E^{--}_R)^T}$, which form an SU$(2)_L$ triplet with hypercharge $Y=-1$.\footnote{These differ from the $Y=0$ triplet in the type III seesaw \cite{type3_seesaw}.} Despite comprising a modest (and testable) extension of the SM, these states have been largely neglected in the literature. Fortunately this deficiency was recently rectified; discovery prospects at the LHC were analyzed in~\cite{DelNobile:2009st}, and their low-energy phenomenology was studied in~\cite{Chua:2010me}. Further studies were subsequently performed by \cite{Delgado:2011iz}. The phenomenology of these exotics was somewhat interesting --- they can induce tree-level  modifications to flavor-changing currents, leading to non-unitary mixing effects in weak interactions, and lepton flavor violation~\cite{Chua:2010me}.  Furthermore, the collider signatures of the doubly-charged lepton $E^{\pm\pm}_R$ are very distinctive~\cite{DelNobile:2009st,Delgado:2011iz}.\footnote{The minimally coupled lepton doublet with $Y=-3/2$ also predicts a doubly charged lepton \cite{DelNobile:2009st, Law:2011qe}, but its phenomenology is markedly different. For related work see~\cite{Kumericki:2011hf}.}

Given that current bounds permit these exotics to be discovered at the LHC~\cite{DelNobile:2009st,Chua:2010me,Delgado:2011iz}, it is worth asking if they could enable new mechanisms for generating neutrino mass. We investigate this matter in the present work, finding that new approaches can arise. In particular, the resulting neutral-lepton mass matrix is suggestive of an inverse seesaw mechanism~\cite{inverse_seesaw}, and we present simple models that achieve a (non-standard) radiative inverse seesaw~\cite{Ma:2009gu}. These models allow one to realize small neutrino masses without invoking tiny parameters, and permit the new physics to appear at the TeV scale. Furthermore, the models contain remnant low-energy symmetries that ensures the stability of a new dark matter candidate --- an unintended but welcome consequence.

The plan of this paper is as follows. We briefly discuss the exotic leptons and elucidate why they are suggestive of an inverse seesaw in Sec.~\ref{sec:nu_mass_with_exotics}. Simple models that achieve a radiative inverse seesaw are presented in Sec.~\ref{sec:radiative_ISS}. Conclusions are drawn in Sec.~\ref{sec:conc} and additional pertinent details appear in an appendix.

\section{Neutrino mass and exotic triplets\label{sec:nu_mass_with_exotics}}

We extend the SM to include the exotic triplets $E_{R,L}$, which transform under
$G_\text{SM}= \text{SU}(3)_c\otimes \text{SU}(2)_L \otimes \text{U}(1)_Y$ as
\beqn{eqn:E_defn}
E_{R,L} = \thoneMat{E^0}{E^-}{E^{--}}_{R,L} \sim (1,3,-1).
\eeqn
The gauge invariant Lagrangian thus includes the terms
\beqn{eqn:E_Lag}
\mathcal{L} \supset - Y_E \overline{L}_L\phi E_R - M_E \overline{E}_L E_R
+\text{H.c.}\;,
\eeqn
where $L_L $ ($\phi$) is the SM lepton (Higgs) doublet. For a given value of $M_E$, the Yukawa couplings $Y_E$ are constrained by $Z$ decays and lepton flavor violation~\cite{Chua:2010me}. We are interested in $M_E\simeq 1$~TeV, in which case the matrix elements must satisfy~\cite{Chua:2010me}
\bea
(Y_E)_{ii} \lesssim 1\;\quad\mathrm{and}\quad(Y_E)_{ij} \lesssim 10^{-2}\ \ \mathrm{for}\ \ i\neq j\;,
\eea
barring tuned cancellations.

With this minimal anomaly-free extension, the neutral lepton mass matrix is
\bea
 \mathcal{M}_\text{inv}^E=
 \ththMat{0}{Y_E \vev{\phi^0}}{0}
         {Y_E^\dagger \vev{\phi^0}}{0}{M_E}
         {0}{M_E}{0} \;,\label{exotic_mass_matrix} 
\eea
in the basis $\{\nu_L,E_R^0,E_L^0\}$. This matrix conserves lepton number symmetry, and one obtains a heavy Dirac neutrino and a massless (mostly SM) neutrino per generation. As a result, additional ingredients are required to generate nonzero SM neutrino masses.\footnote{This is similar to radiative models, for which the extension by one exotic is typically insufficient to generate neutrino mass (see e.g.~\cite{babu_zee,Foot:2007ay}). Also see~\cite{Babu:2009aq}, 
 which includes lepton triplets.} Note, however, that the matrix \eqref{exotic_mass_matrix} is suggestive of two familiar forms in the literature --- namely those of the double seesaw mechanism (see e.g.~\cite{King:2003jb}) and  the inverse seesaw mechanism~\cite{inverse_seesaw}. The former approach is unsuitable in the present context as it requires one to break the electroweak symmetry at a scale $\gg M_E\sim$~TeV. Thus we are led to consider an inverse seesaw. Let us briefly summarize the main features of this approach.

The canonical inverse seesaw  requires one to add a four-component singlet fermion $\psi$  (per family) to the SM.
 Consequently, in the basis $\{\nu_L,\psi_R$,$\psi_L\}$, the mass matrix has the form
\beqn{eqn:inv_basic}
 \mathcal{M}_\text{inv}=
 \ththMat{0}{m}{0}
         {m}{\delta_R}{M}
         {0}{M}{\delta_L} \;,
\eeqn
where $m$ and $M$ are Dirac masses while $\delta_{L,R}$ are (bare) Majorana masses. A field-redefinition freedom ensures that the $\overline{\nu}_L(\psi_L)^c$ entry can be set to zero. When $\delta_{L,R} \neq 0$, lepton number symmetry is explicitly broken and all three fields acquire mass. The inverse seesaw occurs in the technically-natural limit of $\delta_{L,R}\ll m,\,M$, for which the eigenvalues are
\begin{align}
 m_1 &= \frac{m^2\delta_L}{m^2+M^2} + \order{\delta^2} \,, \label{eqn:evals_1}\\
 m_{2,3} &=
 \mp \sqrt{m^2+M^2} - \frac{M^2\delta_L}{2(m^2+M^2)} +\frac{\delta_R}{2}
 + \order{\delta^2}\,.\nonumber
\end{align}
Observe that the light neutrino mass (i.e. $m_1$) is suppressed by the smallness of $\delta_L$, with further suppression if $m \ll M$.

An interesting feature of this scheme is that it allows significant mixing between $\nu_L$ and $\psi_L$, namely $\theta \simeq \mathcal{O}(\sqrt{|m_1/\delta_L|})$. For $m_1 \lesssim$~0.1~eV and $\delta_L \lesssim$~30~keV (compatible with the new physics $M$ being at the TeV scale), this mixing is $\order{10^{-3}}$ and non-unitary mixing effects may be detectable in future experiments~\cite{non_unitary_exp}. Note that the mixing between $\nu_L$ and $\psi_R$ is at the level of $\order{m_1/m} \simeq \order{10^{-9}}$ and remains negligible.\footnote{For these parameters, a normal type-I/III seesaw gives mixing of \order{10^{-7}}. The type-II seesaw \cite{type2_seesaw} does not lead to unitarity violation.}

For our purposes, the key point to note is that when the small parameters vanish ($\delta_{L,R}\rightarrow0$), the mass matrix \eqref{eqn:inv_basic} reduces to that obtained with exotic leptons in \eqref{exotic_mass_matrix}. This suggests that $\delta_{L,R}$ are small because they are zero at tree-level, and that $\delta_{L,R}\ne0$ is generated radiatively by new physics that communicates with $E_{L,R}$. This would, in turn, lead to small SM neutrino masses via a radiative inverse seesaw. In the following section we present simple models that realize this idea.

\section{Radiative inverse seesaw with exotic lepton triplets\label{sec:radiative_ISS}}

We wish to generate small values of $\delta_{L,R}$ to enable an inverse seesaw mechanism involving the triplets $E_{R,L}$. To this end, we consider models that radiatively generate nonzero $\delta_{L,R}$.\footnote{It appears that one could add a scalar $\sigma \sim (1,5,-2)$ to obtain the terms $\delta_{L,R} \overline{E^0_{L,R}} (E^0_{L,R})^c$, with $\delta_{L,R}\propto \vev{\sigma^0}$. However, $\vev{\sigma^0} \neq 0$ produces a Goldstone mode, and tuning is required to obtain $\order{\text{eV}}$ neutrino masses when the new physics is $\mathcal{O}(\mathrm{TeV})$.} We present two such models; one employing an extended gauge symmetry, and another with a global $\mathbb{Z}_4$  symmetry.

\subsection{Model with a gauged U(1)\label{sec:gauge_model}} 

Nonzero $\delta_{L,R}$ can be generated radiatively by introducing new particles that are charged under a hidden (or ``dark'') gauge symmetry. 
We consider the extended gauge group $G_\text{SM}\otimes \text{U}(1)_d$, with new fields that transform as
\bea
& &\quad\quad E_{R,L} \sim (1,3,-1)(0)\;,\;
N_{R,L} \sim (1,1,0)(d)\;,\nonumber\\ 
& &\xi \sim (1,3,-1)(-d),\;\; \eta \sim (1,1,0)(d),\;\; \chi \sim (1,1,0)(2d)\;.\nonumber
\eea
Here the first (second) line contains exotic fermions (scalars), and $d$ is an arbitrary charge that can be set to unity. These particles, together with the SM fields (which are U$(1)_d$ singlets), form an anomaly-free set. In addition to \eqref{eqn:E_Lag}, the Lagrangian contains the terms
\begin{align}
 \mathcal{L}&\supset
  h_\xi \overline{E}_L  \xi N_R
 +h_\chi \overline{N}_R (N_R)^c \chi
 +M_N \overline{N}_L N_R  \nonumber\\
 &\quad
 +h_\xi' \overline{E}_R  \xi N_L
 +h_\chi' \overline{N}_L (N_L)^c \chi+ \text{H.c.}\;,\label{eqn:new_terms_U1d}
\end{align}
where $h_{\xi,\chi}$ and $h_{\xi,\chi}'$ are Yukawa couplings.

Suppose that our model building choices enforce the following VEV pattern
\beqn{eqn:vev_pattern}
\vev{\phi^0} \simeq 174 \text{ GeV}\;,\;\;
\vev{\xi^0},
\vev{\eta}=0\;,\;\;
\vev{\chi} \neq 0\;.
\eeqn
Then, in the basis $\{\nu_L,E_R^0,E_L^0\}$, the tree-level mass matrix has the form \eqref{exotic_mass_matrix} with a conserved lepton-number symmetry. In addition, the $2\times2$ mass matrix for the ``dark'' neutrinos $N_{L,R}$ decouples, and the associated mass eigenstates are two (Majorana) linear combinations of $N_{L,R}$. These dark neutrinos do not mix with the other fields --- a statement that holds to all orders in perturbation theory, as we explain below. 

\begin{figure}[h]
\begin{center}
\includegraphics[width=0.70\columnwidth]{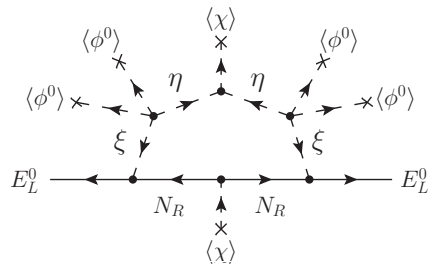}
\caption{One-loop diagram for the Majorana mass $\delta_L$. Note that $\delta_R$ is obtained via $M_E$ mass insertions at both ends.}
\label{fig:1loopdim9}
\end{center}
\end{figure}

The most general scalar potential is
\begin{align}
V_S &=
(\phi^\dagger\phi)\left[\mu_\phi^2 + \lambda_{\phi}(\phi^\dagger\phi)
   + \lambda_{\phi\xi}(\xi^\dagger\xi)
      + \lambda_{\phi\eta}(\eta^\dagger\eta) \right.\nonumber\\
&\;\;\left.
     + \lambda_{\phi\chi}(\chi^\dagger\chi)\right]
 +(\eta^\dagger\eta)\left[\mu_\eta^2 
      + \lambda_{\eta}(\eta^\dagger\eta)
         + \lambda_{\eta\chi}(\chi^\dagger\chi)
\right]\nonumber\\
&\;\;+(\xi^\dagger\xi)\left[\mu_\xi^2 + \lambda_{\xi}(\xi^\dagger\xi)
      + \lambda_{\xi\eta}(\eta^\dagger\eta)
         + \lambda_{\xi\chi}(\chi^\dagger\chi)
\right]\nonumber\\
&\;\;
+(\chi^\dagger\chi)\left[\mu_\chi^2 
         + \lambda_{\chi}(\chi^\dagger\chi)
   \right]\nonumber\\
&\;\;
+\left[\lambda_{\xi\phi\eta}\, \xi\phi\phi\eta
     +\frac{1}{2}\mu_{\eta\chi}\, \eta\eta\chi^\dagger
     + \text{H.c.}\right].\label{eqn:V_S_all}
\end{align}
Observe that lepton number symmetry is broken explicitly in the last line. As a result, one expects the Majorana masses $\delta_{L,R}$ to be generated radiatively. Indeed, $\delta_L$ is generated via the one-loop dim-9 diagram in Fig.~\ref{fig:1loopdim9}. The corresponding diagram for $\delta_R$ is obtained with appropriate mass insertions.
%
%

Writing the scale of new physics as $\Lambda \simeq M_\xi,M_\eta, M_N$ (the masses for $\xi, \eta, N_{R,L}$ respectively),  the one-loop Majorana mass is approximately
\beqn{eqn:deltaL_U1d}
 \delta_L \simeq -\frac{1}{16\pi^2}\,
 \frac{h_\xi^2 h_\chi \lambda_{\xi\phi\eta}^2}{15}\,
\frac{\vev{\phi^0}^4\vev{\chi}^2}{\Lambda^6}\;\mu_{\eta\chi}\;.
\eeqn
For $\vev{\chi} \simeq \Lambda \simeq$~1~TeV, $\mu_{\eta\chi} \simeq$~100~GeV (electroweak scale) and $\lambda_{\xi\phi\eta}, h_{\xi,\chi} =\order{1}$, we have $|\delta_L| \simeq$ \order{10-100}~keV. Using this in (\ref{eqn:evals_1}), with the conservative value $Y_E \simeq 10^{-2}$ for $M_E \simeq 1$~TeV, gives a light neutrino mass of about $10^{-1}$~eV.
Hence, there is no severe tuning and small neutrino masses are compatible with the new physics appearing at the TeV scale.\footnote{The model also generates a three-loop dim-7 diagram for the $\overline{\nu}_L(\nu_L)^c$ mass term. We show in the Appendix that this does not affect our conclusions. There is also a two-loop dim-7 diagram for $\delta_{L,R}$, but this is subdominant when the new states are $\mathcal{O}(\mathrm{TeV})$ (see Sec.~\ref{sec:sec:z4_symmetry}).}

Note that the VEV pattern in  (\ref{eqn:vev_pattern}) is easily achieved. A small hierarchy between \vev{\phi^0} and $\vev{\chi}=\order{\text{TeV}}$ is obtained provided the mixing term $\lambda_{\phi\chi}\phi^\dagger\phi \chi^\dagger\chi$ is not too large. 
The requirement that $\vev{\xi^0},\vev{\eta} =0$ is also compatible with $\vev{\phi^0},\vev{\chi} \neq 0$, as long as all couplings involving $\xi$ and $\eta$ in $V_S$ are positive. 
The stability of the classical vacua may be affected by radiative corrections, including fermion loops induced by $h_\xi, h_\xi'\ne0$. However, for a given set of classical parameters that achieve (\ref{eqn:vev_pattern}), one can always select a suitable cut-off to ensure this minimum persists at the quantum level. If needed, one can also reduce $h_\xi,h_\xi'$ to allow larger values of the cut-off. Smaller values of $h_\xi$ help to suppress $\delta_L$, so smaller $\Lambda$ can be used. 

Since $\xi$ is charged under $G_\text{SM}$, one might naively worry about the SM $\rho$-parameter. However, we require $\vev{\xi^0} =0$, so there are no tree-level corrections to the SM value and the model lies safely within experimental limits. Note that symmetry breaking induces mass mixing between $\chi$ and $\phi^0$, providing a channel for communication between the SM and the hidden sector. There is also mixing between $\xi$ and $\eta$, though these scalars do not mix with $\chi$ and $\phi^0$. 
The hidden sector will also communicate with the SM via gauge kinetic-mixing~\cite{Holdom:1985ag} between U$(1)_d$ and hypercharge. The mixing term in the Lagrangian is
\beqn{kin_mixed_lagrangian}
\mathcal{L} \supset
  -\frac{\kappa}{2}
\left(
   \cos \theta_w F_{\mu\nu}-\sin \theta_w Z_{\mu\nu}
\right)\;Z_d^{\mu\nu}\;,
\eeqn
where $\theta_w$ is the weak mixing angle, and the mixing parameter obeys $\kappa \ll 1$ (see below). To leading order in $\kappa$, the mixed kinetic-Lagrangian is diagonalized by the following field redefinitions:
\begin{align}
A^\mu  &\rightarrow A^\mu - \kappa \cos \theta_w Z_d^\mu\;,\nonumber\\
Z^\mu  &\rightarrow Z^\mu + \kappa \sin \theta_w Z_d^\mu\;,\nonumber\\
Z_d^\mu &\rightarrow Z_d^\mu\;.
\end{align}
In this basis, the mass terms for $Z$ and $Z_d$ are
\beqn{eqn:ZZd_mass_Lag}
\mathcal{L} \supset
 \frac{1}{2}m_Z^2Z^2 + \kappa \sin\theta_w m_Z^2 Z Z_d
 + \frac{1}{2}m_{Z_d}^2 Z_d^2\;.
\eeqn
Hence, to this order, the $Z$-$Z_d$ mixing angle is
\beqn{eqn:ZZd_mixing}
 \theta_{d} \simeq \kappa \sin\theta_w 
 \left(\frac{m_Z^2}{m_Z^2-m_{Z_d}^2}\right) ,
\eeqn
which is suppressed by both small $\kappa$ and the mass ratio $m_Z^2/m_{Z_d}^2$. The corresponding mass eigenvalues are $M_Z^2 \simeq m_Z^2[1+\order{\kappa^2}]$ and
$M_{Z_d}^2 \simeq m_{Z_d}^2[1+\order{\kappa^2}]$. With these results, one obtains the coupling between the \emph{physical} massive vectors and matter:
\bea
 \mathcal{L}\supset
 - Z^{\mu}\left[J_Z+\theta_d J_{Z_d}\right]_\mu 
  - Z_d^\mu \left[
   J_{Z_d} -\kappa J_Y - \theta_{d} J_Z
  \right]_\mu ,\nonumber
\eea
where $J_Y$ is the SM hypercharge current. Thus, the dominant effect of the kinetic mixing is to induce a small coupling of $Z_d$ to $J_Y$ when 
$\kappa \gg \theta_{d}$.

In addition to collider signals due to the triplets $E_{L,R}$, the new vector $Z_d$ can produce observable effects. Detailed phenomenological studies of a hidden vector with mass $M_{Z_d}\gtrsim 200$~GeV have appeared in the literature~\cite{Chang:2006fp}. In the present model, when $Z_d$ and $\chi$ are the lightest exotic states, the constraints on $\kappa$ are similar to those obtained in Ref.~\cite{Chang:2006fp}:
\beqn{eqn:kappa_limit}
 |\kappa| \lesssim 0.006\times \left(\frac{M_{Z_d}}{\text{TeV}}\right)\;.
\eeqn
Clearly, $\kappa\ll1$ is necessary for $M_{Z_d}\lesssim \mathrm{TeV}$, though the constraint is compatible with $\kappa$ being generated at the one-loop level.

Once produced in a collider, the hidden vector can decay back to SM fermions. These decays proceed through the kinetic mixing, and in the limit $M_{Z_d}\gg M_Z$ the widths are 
\beqn{eqn:Zd_partial_width}
 \drate{Z_d\rightarrow \overline{f}f}
 \simeq \frac{\kappa^2 \alpha \,N_c}{6\cos^2 \theta_w} 
 \left[Y_{f_{L}}^2+Y_{f_R}^2\right] M_{Z_d}\;.
\eeqn
Here $\alpha$ is the fine-structure constant, $Y_{f}$ is the hypercharge value, and $N_c$ is the color factor. Notice the distinctive hypercharge dependence,
which ensures that the branching fractions to SM fermions are approximately set by $[Y_{f_{L}}^2+Y_{f_R}^2]$.  Summing over kinematically available final states (including $W^+W^-$) gives~\cite{Chang:2006fp}
\beqn{eqn:Zd_full_width}
\Gamma_{Z_d}\approx 0.2\times
\left(\frac{\kappa^2}{0.01}\right)\;\left(\frac{M_{Z_d}}{\text{TeV}}\right)
\text{GeV}\;.
\eeqn
The $\mathcal{O}(\kappa^2)$ suppression forces the width to be rather narrow, but discovery remains possible in regions of parameter space~\cite{Chang:2006fp}.

Finally, let us point out another welcome feature of the model --- namely the existence of a stable dark matter candidate. 
Note that when $\chi$ develops a nonzero VEV, the U$(1)_d$ symmetry is broken to a residual $\mathbb{Z}_2$. Since $\vev{\xi^0},\vev{\eta} =0$, this symmetry, whose action is defined by 
\beqn{eqn:residual_Z2}
 \{N_{R,L},\xi,\eta\} 
  \rightarrow -\{N_{R,L},\xi,\eta\}\;,\;
 \psi_\text{others} \rightarrow \psi_\text{others}\;,
\eeqn
remains as an exact (accidental) symmetry of the low-energy Lagrangian. This prevents $N_{R,L}$ from mixing with the other leptons, as aforementioned. Furthermore, it forces the lightest particle among $N_{R,L}$, $\xi^0$ and $\eta$ to be absolutely stable, ensuring a dark matter candidate. 
Since neutrino masses can be generated with couplings as large as $\mathcal{O}(1)$, it is trivial to obtain the correct dark matter abundance via freeze-out with $\Lambda \simeq \order{\text{TeV}}$. The dark matter is therefore a standard WIMP whose stability is related to the origin of neutrino mass (see e.g.~\cite{Lindner:2011it}).
The charged components of $\xi$ also transform non-trivially in \eqref{eqn:residual_Z2}. Thus, one must ensure that $\xi^-$ and $\xi^{--}$ can decay in order to avoid having a (cosmologically excluded) stable charged particle. This is easy to achieve. For example, if the dark matter candidate is the lightest linear-combination of $N_{L,R}$, the decay $\xi^-\rightarrow E^-\,N$ is allowed provided $M_\xi> M_E+M_N$. Similarly, if $\chi$ is the dark matter candidate, decays like $\xi^-\rightarrow \chi \;W^-\;\gamma$ can be allowed.\footnote{Treating the $\xi^0$-$\,\eta$ mixing as a mass insertion, the components of $\xi$ are degenerate at the classical level. Loop corrections are expected to lift this degeneracy. If the radiative splittings make the charged components heavier, decays like $\xi^-\rightarrow \xi^0\;W^-\,\gamma$ would also occur (the $W$ could be off-shell).}

Note that the phenomenology of $Z_d$ also depends on the mass spectrum of the exotics. 
For instance, when the dark matter candidate is the lightest linear-combination of $N_{R,L}$, the invisible decay  $Z_d\rightarrow NN$ may occur. The width for this decay is $\Gamma(Z_d\rightarrow NN)\sim g_d^2 M_{Z_d}$, which is expected to dominate the SM width in (\ref{eqn:Zd_partial_width}), given the $\mathcal{O}(\kappa^2)$ suppression of the latter. Consequently, $Z_d$ decays produce missing energy signals in this case. 

\subsection{Model with a global $\mathbb{Z}_4$\label{sec:sec:z4_symmetry}}
We have seen that a simple gauge extension of the SM allows one to generate SM neutrino masses via the exotic leptons $E_{R,L}$. In this subsection we briefly  show that a related radiative inverse seesaw can occur in models with a discrete symmetry. Specifically, we consider a global $\mathbb{Z}_4$ symmetry, with exotic fields that transform under $G_\text{SM} \times \mathbb{Z}_4$ as
\bea
& &\quad\quad E_{R,L} \sim (1,3,-1)(1)\;,\;
N_{R} \sim (1,1,0)(i)\;,\label{eqna:ex_leptons_Z4} \\
& &\xi \sim (1,3,-1)(-i),\;\; \eta \sim (1,1,0)(i),\;\; \chi \sim (1,1,0)(-1)\;.\nonumber
\eea
As before, $E_{R,L}$ and $N_R$ are fermions, while $\xi,\eta$ and $\chi$ are scalars. All SM particles are neutral under $\mathbb{Z}_4$.  Note that the chiral state $N_L$ is no longer required, as the leptons in (\ref{eqna:ex_leptons_Z4}), together with the SM fermions, form an anomaly-free set. As a result, the interaction Lagrangian does not contain the $N_L$-terms shown in (\ref{eqn:new_terms_U1d}) but is otherwise identical. The scalar potential remains as in (\ref{eqn:V_S_all}), with lepton number symmetry broken explicitly, so one expects nonzero $\delta_{L,R}$ to be generated radiatively.

\begin{figure}[ht]
\begin{center}
\includegraphics[width=0.70\columnwidth]{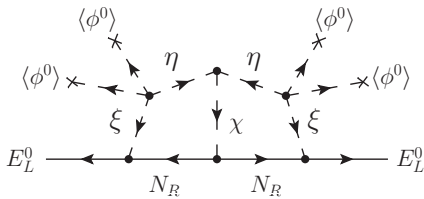}
\caption{Two-loop diagram for the Majorana mass $\delta_L$. $\delta_R$ is obtained via $M_E$ mass insertions at both ends.}
\label{fig:2loopdim7}
\end{center}
\end{figure}

In addition to the change in particle content, we consider the following VEV pattern
\beqn{eqn:VEV_Z4}
 \vev{\phi^0} \simeq 174\ \text{GeV}\;,
 \quad
 \vev{\xi^0},
 \vev{\eta},
 \vev{\chi} =0 \;.
\eeqn
Observe that the $\mathbb{Z}_4$-charged scalars have vanishing VEVs so the discrete symmetry remains intact. This prevents any mixing between $N_R$ and the other leptons. It also forbids the one-loop diagram for $\delta_L$ in Fig.~\ref{fig:1loopdim9}. However, nonzero $\delta_L$ is generated at the two-loop level by the diagram in Fig.~\ref{fig:2loopdim7}. 

In the simplifying limit  $M_\chi\gg\Lambda$, where $\Lambda$ is the (approximate) common mass of the other exotics, one gets\footnote{With  Feynmann parameters one can relate the two-loop integral to that evaluated in~\cite{McDonald:2003zj}.} 
\bea
\delta_{L} \simeq -\frac{1}{(16\pi^2)^2}\  (h_\xi^2 h_\chi  \lambda_{\xi\phi\eta}^2)\ \frac{\langle \phi^0\rangle^4}{M_\chi^2 \Lambda^2}\ \mu_{\chi\eta}.
\eea
Thus, one readily obtains suitable light neutrino masses for natural parameters.

Unsurprisingly, the exact $\mathbb{Z}_4$ symmetry ensures that the lightest $\mathbb{Z}_4$-charged state is a stable dark matter candidate. Distinct from the $U(1)_d$ model, however, one can now have a multi-component dark matter scenario. For example, if $\chi$ is the lightest $\mathbb{Z}_4$-charged state, then the lightest field out of  $\eta$, $\xi$ and $N_R$ is also stable. Similarly, if either of $\eta$, $\xi$ or $N_R$ are the lightest $\mathbb{Z}_4$-charged state, then $\chi$ is also stable if $M_\chi <2 M_\dm$ ($M_\dm$ being the mass of the lightest dark matter candidate). The phenomenology of this model differs from the $U(1)_d$ model due to the absence of $Z_d$ (and $N_L$), but the model is otherwise very similar.

%

\section{Conclusion\label{sec:conc}}
We have presented simple models for neutrino mass that realize a radiative inverse seesaw via the exotic lepton triplets $E_{R,L} \sim (1,3,-1)$. The models employ an extended field content, and allow multiple new states to appear at the TeV scale. The leptons $E_{R,L}$, and the hidden vector $Z_d$ (appearing in one model), are good candidates for new physics that can probed at the LHC. Furthermore, the models contain stable dark matter candidates due to remnant low-energy symmetries.

%

\begin{acknowledgments}
The authors thank C.-H.~Chen and R.~R.~Volkas for useful discussions. 
S.S.C.L is supported in part by the NSC under Grant No. NSC-100-2811-M-006-019 and in part by the NCTS of Taiwan.
\end{acknowledgments}

\appendix

\section{}
Adding a nonzero $\overline{\nu}_L(\nu_L)^c$ entry to the inverse seesaw mass matrix~\eqref{eqn:inv_basic} gives
\beqn{eqn:new_inv_basic}
  \mathcal{M}_\text{inv}=
 \ththMat{\delta_m}{m}{0}
         {m}{\delta_R}{M}
         {0}{M}{\delta_L}.
\eeqn
In the small $\delta_m$ limit the eigenvalues are now given by
\begin{align}
 m_1 &= \underbrace{\frac{M^2\delta_m}{m^2+M^2}}_{\simeq \,\delta_m}+ \frac{m^2\delta_L}{m^2+M^2} + \order{\delta^2} 
     \;,
      \label{eqn:new_evals_1}\\
  m_{2,3}    &=
 \mp \sqrt{m^2+M^2} +\frac{m^2\delta_m}{2(m^2+M^2)}
 - \frac{M^2\delta_L}{2(m^2+M^2)} \nonumber\\
 &\quad
 +\frac{\delta_R}{2}
 + \order{\delta^2}\,.\label{eqn:new_evals_23}
\end{align}
Thus, provided $\delta_m \not \gg m^2\delta_L/(m^2+M^2)$, the lightest eigenvalue remains on the order of the standard inverse seesaw result, $m_1=\mathcal{O}(m^2 \delta_L/M^2)$.

\begin{figure}[t]
\begin{center}
\includegraphics[width=0.80\columnwidth]{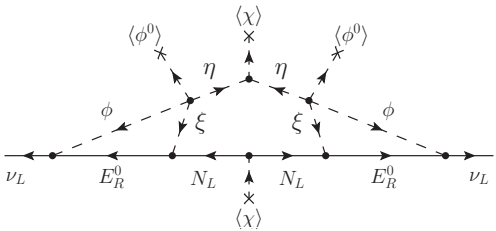}
\caption{A three-loop dim-7 contribution to the $\overline{\nu}_L(\nu_L)^c$ entry of the neutral lepton mass matrix in the U$(1)_d$ model. }
\label{fig:3loopnumass2}
\end{center}
\end{figure}

In the $U(1)_d$ model of Sec.~\ref{sec:gauge_model}, nonzero $\delta_m$ is generated at the three-loop level by the diagram shown in Fig.~\ref{fig:3loopnumass2}. 
In the large $M_\xi$ limit one can approximate this diagram as
%
%

%
\bea
\delta_m&\simeq&   \frac{8\, h_\xi^2 h_\chi \lambda_{\xi\phi\eta}^2}{(16\pi^2)^3}\ \frac{Y_E^2\langle \phi^0\rangle^2\langle\chi\rangle^2}{3M_\xi^4}\ \mu_{\chi\eta}\nonumber\\
& &\qquad\qquad \times \left\{\log\left[\frac{M^2_\xi+\Lambda^2}{\Lambda^2}\right]-1\right\}^2,\label{3_loop_mass}
\eea
where $\Lambda \,(\,\gg m_{\phi^0})$ generically denotes the mass of the other exotics. When the exotics are at the TeV scale, the result \eqref{3_loop_mass} is subdominant to the one-loop expression obtained in Sec.~\ref{sec:gauge_model}, due to the additional loop-factor suppression.


\begin{thebibliography}{99}
  
  
\bibitem{type1_seesaw}
  P.~Minkowski,
  Phys.\ Lett.\ B {\bf 67}, 421 (1977);
%
  T.~Yanagida, in {\it Workshop on Unified Theories}, KEK report 79-18 p.95 (1979);
%
  M.~Gell-Mann, P.~Ramond, R.~Slansky,
  in {\it Supergravity} (North Holland, Amsterdam, 1979)
  eds. P.~van~Nieuwenhuizen, D.~Freedman, p.315.;
%
  S.~L.~Glashow, in {\it 1979 Cargese Summer Institute on Quarks and Leptons} (Plenum Press,
  New York, 1980) eds. M.~Levy, J.-L.~Basdevant, D.~Speiser, J.~Weyers, R.~Gastmans and
  M.~Jacobs, p.687;
%
  R.~Barbieri, D.~V.~Nanopoulos, G.~Morchio and F.~Strocchi,
  Phys.\ Lett.\ B {\bf 90}, 91 (1980);
%
  R.~N.~Mohapatra and G.~Senjanovic,
  Phys.\ Rev.\ Lett.\  {\bf 44}, 912 (1980).


\bibitem{Langacker:2011bi} 
  P.~Langacker,
  arXiv:1112.5992 [hep-ph].


\bibitem{ArkaniHamed:1998pf} 
  N.~Arkani-Hamed and Y.~Grossman,
  Phys.\ Lett.\ B {\bf 459}, 179 (1999)
  [hep-ph/9806223];
  Y.~Grossman and D.~J.~Robinson,
  JHEP {\bf 1101}, 132 (2011)
  [arXiv:1009.2781 [hep-ph]].



\bibitem{McDonald:2010jm} 
  K.~L.~McDonald,
  Phys.\ Lett.\ B {\bf 696}, 266 (2011)
  [arXiv:1010.2659 [hep-ph]].

\bibitem{Duerr:2011ks} 
  M.~Duerr, D.~P.~George and K.~L.~McDonald,
  JHEP {\bf 1107}, 103 (2011)
  [arXiv:1105.0593 [hep-ph]].

\bibitem{Okui:2004xn} 
  T.~Okui,
  JHEP {\bf 0509}, 017 (2005)
  [hep-ph/0405083].

\bibitem{Hundi:2011et} 
  R.~S.~Hundi and S.~Roy,
  Phys.\ Lett.\ B {\bf 702}, 228 (2011)
  [arXiv:1105.0291 [hep-ph]].

\bibitem{babu_zee}
  A.~Zee,
  Nucl.\ Phys.\  B {\bf 264}, 99 (1986);
%
  K.~S.~Babu,
  Phys.\ Lett.\  B {\bf 203}, 132 (1988).


\bibitem{Chen:2011de} 
  M.~-C.~Chen and J.~Huang,
  Mod.\ Phys.\ Lett.\ A {\bf 26}, 1147 (2011)
  [arXiv:1105.3188 [hep-ph]].


\bibitem{DelNobile:2009st}
  E.~Del Nobile, R.~Franceschini, D.~Pappadopulo and A.~Strumia,
  Nucl.\ Phys.\  B {\bf 826}, 217 (2010)
  [arXiv:0908.1567 [hep-ph]].

\bibitem{Chua:2010me}
  C.~K.~Chua and S.~S.~C.~Law,
  Phys.\ Rev.\  D {\bf 83}, 055010 (2011)
  [arXiv:1011.4730 [hep-ph]].


\bibitem{type3_seesaw}
  R.~Foot, H.~Lew, X.~G.~He and G.~C.~Joshi,
  Z.\ Phys.\  C {\bf 44}, 441 (1989),


\bibitem{Delgado:2011iz} 
  A.~Delgado, C.~Garcia Cely, T.~Han and Z.~Wang,
  Phys.\ Rev.\ D {\bf 84}, 073007 (2011)
  [arXiv:1105.5417 [hep-ph]].

\bibitem{Law:2011qe} 
  S.~S.~C.~Law,
  JHEP {\bf 1202}, 127 (2012)
  [arXiv:1106.0375 [hep-ph]].

\bibitem{Kumericki:2011hf} 
  K.~Kumericki, I.~Picek and B.~Radovcic,
  Phys.\ Rev.\ D {\bf 84}, 093002 (2011)
  [arXiv:1106.1069 [hep-ph]];
  S.~Biondini, O.~Panella, G.~Pancheri, Y.~N.~Srivstava and L.~Fano,
  arXiv:1201.3764 [hep-ph].


\bibitem{inverse_seesaw}
  D.~Wyler and L.~Wolfenstein,
  Nucl.\ Phys.\ B {\bf 218}, 205 (1983);
  %
  R.~N.~Mohapatra and J.~W.~F.~Valle,
  Phys.\ Rev.\ D {\bf 34}, 1642 (1986);
%
  E.~Ma,
  Phys.\ Lett.\ B {\bf 191}, 287 (1987).

\bibitem{Ma:2009gu} 
  E.~Ma,
  Phys.\ Rev.\ D {\bf 80}, 013013 (2009)
  [arXiv:0904.4450 [hep-ph]].
  Also see 
  F.~Bazzocchi,
  Phys.\ Rev.\ D {\bf 83}, 093009 (2011)
  [arXiv:1011.6299 [hep-ph]].


\bibitem{Foot:2007ay} 
  R.~Foot, A.~Kobakhidze, K.~L.~McDonald and R.~R.~Volkas,
  Phys.\ Rev.\ D {\bf 76}, 075014 (2007)
  [arXiv:0706.1829 [hep-ph]].


\bibitem{Babu:2009aq} 
  K.~S.~Babu, S.~Nandi and Z.~Tavartkiladze,
  Phys.\ Rev.\ D {\bf 80}, 071702 (2009)
  [arXiv:0905.2710 [hep-ph]];
%
  E.~Ma,
  Mod.\ Phys.\ Lett.\ A {\bf 24}, 2491 (2009)
  [arXiv:0905.2972 [hep-ph]];
  %
  D.~Ibanez, S.~Morisi and J.~W.~F.~Valle,
  Phys.\ Rev.\ D {\bf 80}, 053015 (2009)
  [arXiv:0907.3109 [hep-ph]].
%




\bibitem{King:2003jb} 
  S.~F.~King,
  Rept.\ Prog.\ Phys.\  {\bf 67}, 107 (2004)
  [hep-ph/0310204].
%


\bibitem{non_unitary_exp}
  V.~Barger, S.~Geer and K.~Whisnant,
  New J.\ Phys.\  {\bf 6}, 135 (2004)
  [hep-ph/0407140];
%
  S.~Antusch, M.~Blennow, E.~Fernandez-Martinez and J.~Lopez-Pavon,
  Phys.\ Rev.\ D {\bf 80}, 033002 (2009)
  [arXiv:0903.3986 [hep-ph]];
%
  M.~Malinsky, T.~Ohlsson and H.~Zhang,
  Phys.\ Rev.\ D {\bf 79}, 073009 (2009)
  [arXiv:0903.1961 [hep-ph]];
%
  Z.~-z.~Xing and S.~Zhou,
  Phys.\ Lett.\ B {\bf 666}, 166 (2008)
  [arXiv:0804.3512 [hep-ph]];
%
  S.~Goswami and T.~Ota,
  Phys.\ Rev.\ D {\bf 78}, 033012 (2008)
  [arXiv:0802.1434 [hep-ph]];
%
  M.~Malinsky, T.~Ohlsson, Z.~-z.~Xing and H.~Zhang,
  Phys.\ Lett.\ B {\bf 679}, 242 (2009)
  [arXiv:0905.2889 [hep-ph]].
  

\bibitem{type2_seesaw}
  G.~Senjanovic and R.~N.~Mohapatra,
  Phys.\ Rev.\  D {\bf 12}, 1502 (1975);
%
  W.~Konetschny and W.~Kummer,
  Phys.\ Lett.\  B {\bf 70}, 433 (1977);
%
  T.~P.~Cheng and L.~F.~Li,
  Phys.\ Rev.\  D {\bf 22}, 2860 (1980);
%
  M.~Magg and C.~Wetterich,
  Phys.\ Lett.\  B {\bf 94}, 61 (1980);
%
  J.~Schechter and J.~W.~F.~Valle,
  Phys.\ Rev.\  D {\bf 22}, 2227 (1980);
%
  G.~Lazarides, Q.~Shafi and C.~Wetterich,
  Nucl.\ Phys.\  B {\bf 181}, 287 (1981);
%
  C.~Wetterich,
  Nucl.\ Phys.\  B {\bf 187}, 343 (1981);
%
  R.~N.~Mohapatra and G.~Senjanovic,
  Phys.\ Rev.\  D {\bf 23}, 165 (1981).

\bibitem{Holdom:1985ag}
  B.~Holdom,
  Phys.\ Lett.\  B {\bf 166}, 196 (1986);
  R.~Foot and X.~G.~He,
  Phys.\ Lett.\  B {\bf 267}, 509 (1991).

\bibitem{Chang:2006fp} 
  W.~-F.~Chang, J.~N.~Ng and J.~M.~S.~Wu,
  Phys.\ Rev.\ D {\bf 74}, 095005 (2006)
  [Erratum-ibid.\ D {\bf 79}, 039902 (2009)]
  [hep-ph/0608068].
 
\bibitem{Lindner:2011it} 
  M.~Lindner, D.~Schmidt and T.~Schwetz,
  Phys.\ Lett.\ B {\bf 705}, 324 (2011)
  [arXiv:1105.4626 [hep-ph]];
  F.~-X.~Josse-Michaux and E.~Molinaro,
  Phys.\ Rev.\ D {\bf 84}, 125021 (2011)
  [arXiv:1108.0482 [hep-ph]];
  S.~Kanemura, T.~Nabeshima and H.~Sugiyama,
  Phys.\ Rev.\ D {\bf 85}, 033004 (2012)
  [arXiv:1111.0599 [hep-ph]].

\bibitem{McDonald:2003zj} 
  K.~L.~McDonald and B.~H.~J.~McKellar,
  hep-ph/0309270.

\end{thebibliography}
\end{document}